# Weighted least squares methods for prediction in the functional data linear model

Aurore Delaigle
Department of Mathematics, University of Bristol, Bristol BS8 1TW, UK and
Department of Mathematics and Statistics, University of Melbourne, VIC, 3010,
Australia

Peter Hall
Department of Mathematics and Statistics, University of Melbourne, VIC, 3010,
Australia and Department of Statistics, University of California at Davis, Davis,
CA 95616, USA

Tatiyana V. Apanasovich
Thomas Jefferson University, Department of Pharmacology and Experimental
Therapeutics, Philadelphia, PA 1910, USA

**Abstract**

The problem of prediction in functional linear regression is conventionally addressed by reducing dimension via the standard principal component basis. In this paper we show that an alternative basis chosen through weighted least-squares, or weighted least-squares itself, can be more effective when the experimental errors are heteroscedastic. We give a concise theoretical result which demonstrates the effectiveness of this approach, even when the model for the variance is inaccurate, and we explore the numerical properties of the method. We show too that the advantages of the suggested adaptive techniques are not found only in low-dimensional aspects of the problem; rather, they accrue almost equally among all dimensions.

**Keywords**: Cross-validation; eigenfunction; eigenvector; functional data analysis; functional linear regression; mean squared error; orthogonal series; principal component analysis; rate of convergence; weighted linear regression.

# 1 Introduction

The functional linear model has the appearance of being rather conventional. It involves representing a scalar response, $Y$, as

$$Y = \alpha + \int_{\mathcal{I}} \beta\, X + \text{error}, \tag{1}$$

where $X$ denotes the function-valued explanatory variable, $\alpha$ is a scalar, $\beta$ is the function-valued slope parameter, and $\mathcal{I}$ is a known compact interval. However, estimation of $\beta$ is generally a nonparametric problem, and the level of complexity implicit in that property can carry over to the problem of prediction, in which we wish to estimate $\alpha + \int_{\mathcal{I}} \beta\, x$ for a given function $x$. Sometimes $\alpha + \int_{\mathcal{I}} \beta\, x$ can be estimated root-$n$ consistently, where $n$ denotes sample size, but more commonly, estimators converge at strictly slower rates. Cai and Hall (2006) discuss these issues, and Faraway (1997), Ferraty and Vieu (2000), Cuevas *et al.* (2002), Ramsay and Silverman (2005, Chapter 12), Cardot *et al.* (2006) and Cardot and Sarda (2006) address functional linear regression in more general terms.

A standard approach to estimating $\alpha$ and $\beta$ is to first estimate the principal component basis from a sample of observations of $(X,Y)$, and then construct an estimator of $\mu(x) = \alpha + \int_{\mathcal{I}} \beta\, x$ in terms of that basis, using least squares. However, in practice the distribution of the error in (1) can be a functional of the distribution of $X$, and the optimal choice of basis can depend significantly on $x$. To address these challenges we could construct the basis so that it gave greater emphasis to observations of $X$ that were relatively close to $x$. For example, we could restrict attention to $X$ for which $\|X - x\| \leq \delta$, where $\|\cdot\|$ was a suitable distance measure and $\delta$ played the role of bandwidth, although $\delta$ would not necessarily be chosen to converge to zero as $n$ increased. More subtly, the basis could be constructed by applying kernel weights to each observation. See Mas (2008) for theoretical results addressing problems of this type.

Although this approach is attractive, practical difficulties can arise from the implicit reduction in sample size that is involved. An alternative method is to estimate the variance, $\sigma(X)^2$ say, of the distribution of the error in (1) conditional on $X$, and adapt prediction to the level of variability there. We suggest solving this problem by modelling $\sigma(x)^2$ as a function of $\alpha + \int_{\mathcal{I}} \beta\, x$, and using its inverse, with $x$ replaced by a data value $X$, as a weight in the basic least-squares problem. We then show that calculations can be simplified by computing a new principal component basis, adapted to heteroscedasticity. While our approach has some similarities with the weighted least squares method used for finite dimensional data, it differs significantly due to the intrinsic nonparametric, and infinite dimensional, characters of functional linear regression; we quantify these issues in theoretical terms.

In summary, this paper makes three main contributions. First, we show in section 2 that adaptive modification of the standard principal component basis, or a nearly-equivalent method based on weighted least-squares, can be advantageous when undertaking functional linear prediction, i.e. when estimating $\mu(x)$. Secondly, we suggest approximations to the value of $\sigma(x)^2$, and we employ them to construct a second basis, this time adapted to heteroscedasticity. Then, in sections 3 and 4 we show that this approach can give real and effective reductions in mean squared error, even when the model we use to estimate variance is not completely correct. An alternative approach would be to use a nonparametric method to estimate variance. However, unsurprisingly given the high degree of noise associated with estimating nonparametric functions, numerical work shows that parametric methods are preferable. These results all have analogues in cases where $Y$ is a multivariate response, although for simplicity and transparency we focus only on the univariate case.

The main theoretical result in section 3 gives a concise account of the way adaptive methods can improve the performance of estimators in functional linear regression. In particular, we show that the advantages accrue almost equally among all dimensions; they are not principally to be found in low-dimensional aspects of the problem.

Previous developments of principal components analysis for functional data play a central role in our work. Early contributions include those of Besse and Ramsay (1986), Ramsay and Dalzell (1991) and Rice and Silverman (1991). From that point a very substantial literature has developed, including but by no means limited to the work of Silverman (1995, 1996), Brumback and Rice (1998), Cardot *et al.* (1999, 2000, 2003), Cardot (2000), Girard (2000), James *et al.* (2000), Boente and Fraiman (2002), He, Müller and Wang (2003), Ramsay and Silverman (2005, Chapters 8–10), Yao *et al.* (2005), Hall and Hosseini-Nasab (2006), Jank and Shmueli (2006), Ocaña *et al.* (2007), Reiss and Ogden (2007) and Huang *et al.* (2008).

## 2 Methodology

### 2.1 Orthogonal series approach to inference in the linear model

The functional linear model argues that independent data pairs $(X_{[1]}, Y_1), \ldots, (X_{[n]}, Y_n)$, distributed as $(X, Y)$, are generated as

$$Y = \alpha + \int_{\mathcal{I}} \beta X + \epsilon, \quad (2)$$

where $\alpha$ is a scalar, $\beta$ and $X$ are functions defined on the compact interval $\mathcal{I}$, and $E(\epsilon \,|\, X) = 0$. Square-bracketed subscripts here distinguish the $i$th observation of $X$, $X_{[i]}$, from the $i$th principal component score, which is conventionally represented by $X_i$. The prediction problem is that of estimating $\mu(x) = E(Y \,|\, X = x) = \alpha + \int_{\mathcal{I}} \beta\, x$ with $(\alpha, \beta)$ at (2), where $x$ denotes a particular value of $X$ and $\mu$ is a scalar functional.

A standard approach to estimating $\mu(x)$ is to introduce an orthonormal basis, say $\psi_1, \psi_2, \ldots$, and argue that $\beta$ and $x$ admit convergent expansions with respect to this sequence, i.e.

$$\beta = \sum_{j=1}^{\infty} b_j \, \psi_j, \quad x = \sum_{j=1}^{\infty} x_j \, \psi_j, \quad \mu(x) = \alpha + \sum_{j=1}^{\infty} b_j \, x_j, \quad (3)$$

where $b_j = \int_{\mathcal{I}} \beta \, \psi_j$ and $x_j = \int_{\mathcal{I}} x \, \psi_j$. Estimators $\hat{\alpha}$ of $\alpha$ and $\hat{b}_j$ of $b_j$, for $j \geq 1$, are then constructed from the data by minimising

$$S_r(\alpha, b_1, \ldots, b_r) = \sum_{i=1}^{n} \left( Y_i - \alpha - \sum_{j=1}^{r} b_j \, X_{ij} \right)^2, \quad (4)$$

where $X_{ij} = \int X_{[i]} \, \psi_j$ and $r$ denotes the frequency cut-off, a smoothing parameter. These definitions of $\hat{\alpha}$ and $\hat{b}_1, \ldots, \hat{b}_r$ reflect the definitions of $\alpha$ and $\beta$ at (2) and, for appropriate choice of $r$, ensure consistency. The resulting estimator of $\mu$ is

$$\hat{\mu}(x) = \hat{\alpha} + \sum_{j=1}^{r} \hat{b}_j \, x_j. \quad (5)$$

A thresholding method could also be used instead of "cut-off smoothing," but the difficulty of estimating the variance of $\hat{b}_j$ makes that approach unattractive.

## 2.2 Principal component basis

It is common to take $\psi_1, \psi_2, \ldots$ to be the principal component basis, ordered so that the corresponding eigenvalues form a decreasing sequence. Specifically, define $K(s,t) = \text{cov}\{X(s), X(t)\}$ to be the covariance function of $X$, and construct the spectral decomposition of $K$,

$$K(s,t) = \sum_{j=1}^{\infty} \theta_j \, \psi_j(s) \, \psi_j(t), \tag{6}$$

where $\theta_1 \geq \theta_2 \geq \ldots \geq 0$ and $(\theta_j, \psi_j)$ are the (eigenvalue, eigenfunction) pairs of the transformation that takes $\psi$ to $K\psi$, defined by $(K\psi)(t) = \int_{\mathcal{I}} K(s,t) \, \psi(s) \, ds$. Then the orthonormal functions $\psi_j$ make up the principal component basis. The $j$th uncentred principal component score of $X$ is $X_j = \int_{\mathcal{I}} X \, \psi_j$.

In practice the principal component basis is unknown, and needs to be estimated from data. To this end we define

$$\widehat{K}(s,t) = n^{-1} \sum_{i=1}^{n} \{X_{[i]}(s) - \bar{X}(s)\} \{X_{[i]}(t) - \bar{X}(t)\} = \sum_{j=1}^{\infty} \hat{\theta}_j \, \hat{\psi}_j(s) \, \hat{\psi}_j(t),$$

where $\bar{X} = n^{-1} \sum_i X_{[i]}$, $\widehat{K}(s,t)$ is an estimator of $K(s,t)$, $(\hat{\theta}_j, \hat{\psi}_j)$ are (eigenvalue, eigenfunction) pairs for the transformation represented by $\widehat{K}$, and the order of the indices $j$ is chosen to ensure that $\hat{\theta}_1 \geq \hat{\theta}_2 \geq \ldots$ Then $\hat{\theta}_j$ and $\hat{\psi}_j$ are our estimators of $\theta_j$ and $\psi_j$, respectively, and we would replace (4) by

$$\hat{S}_r(\alpha, b_1, \ldots, b_r) = \sum_{i=1}^{n} \left( Y_i - \alpha - \sum_{j=1}^{r} b_j \, \widehat{X}_{ij} \right)^2, \tag{7}$$

where $\widehat{X}_{ij} = \int_{\mathcal{I}} X_{[i]} \, \hat{\psi}_j$, giving the obvious estimator $\hat{\mu}(x)$ of $\mu(x)$. Equivalently, since $\hat{\alpha} = \bar{Y} - \int_{\mathcal{I}} \hat{\beta} \, \bar{X}$ then, writing $\overline{\widehat{X}}_j = n^{-1} \sum_i \widehat{X}_{ij}$, we can minimise

$$\hat{S}_r^{\text{equiv}}(b_1, \ldots, b_r) = \sum_{i=1}^{n} \left\{ Y_i - \bar{Y} - \sum_{j=1}^{r} b_j \left( \widehat{X}_{ij} - \overline{\widehat{X}}_j \right) \right\}^2 \tag{8}$$

over $b_1, \ldots, b_r$, obtaining the same numerical values $\hat{b}_1, \ldots, \hat{b}_r$ as we do when minimising (7). Then, defining $\hat{x}_j = \int_{\mathcal{I}} x \, \hat{\psi}_j$, we take

$$\hat{\mu}(x) = \bar{Y} + \sum_{j=1}^{r} \hat{b}_j \left( \hat{x}_j - \overline{\widehat{X}}_j \right) \tag{9}$$

to be our estimator of $\mu(x)$. In a slight abuse of notation, when discussing practical implementation we shall write $\hat{\alpha}$ and $\hat{b}_j$ for the quantities that minimise (7) rather than (4).

## 2.3 Adapting to the variance of $Y$

The estimator $\hat{\mu}$ at (9) is conventional, but does not take into account the fact that the errors at (2) are often heteroscedastic. When a significant amount of variability is explained by that aspect of the problem, we should replace $\hat{S}_r(\alpha, b_1, \ldots, b_r)$ at (7) by its form where a weight, equal to an approximation to the inverse of the variance of $Y_i - \alpha - \int_{\mathcal{I}} \beta\, X_{[i]}$ conditional on $X_{[i]}$, is incorporated into the series at (7).

In conventional parametric regression, the conditional variance of the regression errors is often modelled as a power of the assumed parametric form of $E(Y|X)$. See for example Carroll and Ruppert (1988). In the functional data context we propose modeling $\mathrm{var}(\epsilon\,|\,X)$ by

$$\sigma(X)^2 = g\left(\alpha + \sum_{j=1}^{r} b_j\, X_j\right), \tag{10}$$

with $g$ a univariate function and $\alpha$, $b_1, \ldots, b_r$ and $r$ as in (7) and where $X_j$ is the $j$th principal component score. An adaptive form of $g$ that is often suitable is the "power of the mean" model

$$g(u) = |c_1 u|^{c_2}, \tag{11}$$

where $c_1$ and $c_2$ are constants, or the version of it which includes an intercept term. See Carroll and Ruppert (1988, pp. 5, 65).

In principle, a nonparametric estimator of $g$ could be used. However, in applications to real functional data, where sample sizes are generally only moderate in size, we found that the increase in variance resulting from the nonparametric fit significantly outweighed any improvements in performance that might be expected using that approach. Bear in mind that it is not necessary to have consistent estimators of the variances in order to enjoy improved statistical performance, even in the asymptotic limit. We shall take this point up again in section 3; see point (ii) below the Theorem there.

To estimate $\sigma(X)^2$ we interpret the unweighted estimators $\hat{\alpha}$ and $\hat{\beta} = \sum_{j \leq r} \hat{b}_j\, \hat{\psi}_j$ as pilot estimators of $\alpha$ and $\beta = \sum_j b_j\, \psi_j$, respectively, and use them to calculate residuals $\hat{\epsilon}_i = Y_i - \hat{\alpha} - \sum_j \hat{b}_j\, \widehat{X}_{ij}$. Since these quantities are already centred, we define

$$\widehat{T}(c_1, c_2) = \sum_{i=1}^{n} \left\{ \hat{\epsilon}_i^2 - \left| c_1\left(\hat{\alpha} + \sum_{j=1}^{r} \hat{b}_j\, \widehat{X}_{ij}\right) \right|^{c_2} \right\}^2 \tag{12}$$

and choose $\hat{c}_1$ and $\hat{c}_2$ to minimise $\widehat{T}(c_1, c_2)$. In this notation our estimator of $\mathrm{var}(Y - \alpha - \int_{\mathcal{I}} \beta\, X \mid X = x)$ is, when $x = X_{[i]}$,

$$\widehat{w}(X_{[i]})^{-1} = \left| \hat{c}_1 \left( \hat{\alpha} + \sum_{j=1}^{r} \hat{b}_j\, \widehat{X}_{ij} \right) \right|^{\hat{c}_2}.$$

Next we incorporate these weights into the objective function at (8), obtaining:

$$\widehat{U}_t(b_1, \ldots, b_t) = \sum_{i=1}^{n} \left\{ Y_i - \bar{Y}_w - \sum_{j=1}^{t} b_j \left( \widehat{X}_{ij} - \overline{\widehat{X}}_{j,w} \right) \right\}^2 \widehat{w}(X_{[i]}), \quad (13)$$

where $\bar{Y}_w = \{\sum_i \widehat{w}(X_{[i]})\}^{-1} \sum_i \widehat{w}(X_{[i]})\, Y_i$ and $\overline{\widehat{X}}_{j,w} = \{\sum_i \widehat{w}(X_{[i]})\}^{-1} \sum_i \widehat{w}(X_{[i]})\, \widehat{X}_{ij}$; and we choose $\tilde{b}_{w1}, \ldots, \tilde{b}_{wt}$ to minimise $\widehat{U}_t(b_1, \ldots, b_t)$. A new estimator of $\mu(x)$ is given by the following analogue of (5), based on the new coefficient estimators:

$$\widetilde{\mu}_w(x) = \bar{Y}_w + \sum_{j=1}^{t} \tilde{b}_{wj} \left( \hat{x}_j - \overline{\widehat{X}}_{j,w} \right). \quad (14)$$

A computational advantage of defining estimators by minimising $\widehat{S}_r(\alpha, b_1, \ldots, b_r)$ at (4), rather than $\widehat{U}_t(b_1, \ldots, b_t)$ at (13), is that the "ex transpose ex" matrix in the former case is simple to invert. Indeed, by definition of $\widehat{X}_{ij}$ in terms of the orthogonal functions $\hat{\psi}_j$, the matrix with $(j,k)$th term $n^{-1} \sum_i (\widehat{X}_{ij} - \overline{\widehat{X}}_j)(\widehat{X}_{ik} - \overline{\widehat{X}}_k)$ is diagonal. The fact that this does not hold in the case of the objective function $\widehat{U}_t(b_1, \ldots, b_t)$ reflects the fact that the orthonormal basis functions $\hat{\psi}_j$ are not necessarily, in this case, the natural ones. Instead we could replace $\widehat{U}_t(b_1, \ldots, b_t)$ by

$$\widehat{V}_s(b_1, \ldots, b_s) = \sum_{i=1}^{n} \left\{ Y_i - \bar{Y}_w - \sum_{j=1}^{s} b_j \left( \check{X}_{ij} - \overline{\check{X}}_{j,w} \right) \right\}^2 \widehat{w}(X_{[i]}), \quad (15)$$

where we define $\check{X}_{ij} = \int_{\mathcal{I}} X_{[i]}\, \hat{\phi}_j$ and $\overline{\check{X}}_{j,w} = \{\sum_i \widehat{w}(X_{[i]})\}^{-1} \sum_i \widehat{w}(X_{[i]})\, \check{X}_{ij}$, and where the orthonormal functions $\hat{\phi}_1, \hat{\phi}_2, \ldots$, with corresponding eigenvalues $\widehat{\omega}_1 \geq \widehat{\omega}_2 \geq \ldots$, are defined by the following spectral decomposition

$$\left\{ \sum_{i=1}^{n} \widehat{w}(X_{[i]}) \right\}^{-1} \sum_{i=1}^{n} \{ X_{[i]}(s) - \bar{X}_w(s) \} \{ X_{[i]}(t) - \bar{X}_w(t) \}\, \widehat{w}(X_{[i]}) = \sum_{j=1}^{\infty} \widehat{\omega}_j\, \hat{\phi}_j(s)\, \hat{\phi}_j(t).$$

Taking $\check{b}_{w1}, \ldots, \check{b}_{ws}$ to minimise $\widehat{V}_s(b_1, \ldots, b_s)$, a competitor with $\widetilde{\mu}_w(x)$ at (14) is given by

$$\check{\mu}_w(x) = \bar{Y}_w + \sum_{j=1}^{s} \check{b}_{wj} \left( \int_{\mathcal{I}} x\, \hat{\phi}_j - \overline{\check{X}}_{j,w} \right). \quad (16)$$

The numerical differences between $\widehat{\mu}_w$ and $\check{\mu}_w$ are generally very small.

## 2.4 Practical choice of smoothing parameters

The methodology outlined in sections 2.2 and 2.3 involves two smoothing parameters: $r$, in the equivalent objective functions $\hat{S}_r$ and $\hat{S}_r^{\text{equiv}}$ at (7) and (8), and $t$, in $\widehat{U}_t$ at (13), or $s$, in $\widehat{V}_s$ at (15). We propose selecting these parameters by cross-validation, as follows. Omit the data pair $(X_{[i]}, Y_i)$ from the sample, and, using the remaining $n - 1$ pairs, construct the predictor $\check{\mu}_w(x)$ at (16) for a general $r$ and $s$; denote it by $\check{\mu}_{w,-i}(x \,|\, r, s)$. Put $W(r,s) = \sum_i \{Y_i - \check{\mu}_{w,-i}(X_{[i]} \,|\, r, s)\}^2$, and choose $(r, s)$ to minimise $W(r, s)$. The same approach is used to select $r$ and $t$ for the predictor $\widetilde{\mu}_w(x)$ at (14).

## 3 Theoretical properties

We shall suppose that we model the variance $\text{var}(\epsilon \,|\, X) = \sigma(X)^2$ as $\tau(X)^2$, where the function $\tau$ is known but may not equal $\sigma$. That is, our model may not actually be correct. We shall make three simplifying assumptions: (a) The principal components $\int_{\mathcal{I}} X \, \psi_j$ of $X$ are independent, rather than merely uncorrelated; (b) the principal component basis $\psi_1, \psi_2, \ldots$ is known; and (c):

$\epsilon = \sigma(X)\,\delta$, where $\delta$ is stochastically independent of $X$, $E(\delta) = 0$, $E(\delta^2) = 1$, the functional $\sigma$ is bounded, $\tau$ is bounded above zero, and $\sigma(X)$ and $\tau(X)$ depend on only a finite number of the principal component scores $X_j = \int_{\mathcal{I}} X \, \psi_j$; that is, for some $t \geq 1$ and positive integers $j_1, \ldots, j_t$ we can write $\sigma(X)^2 = \text{var}(\epsilon \,|\, X) = h(X_{j_1}, \ldots, X_{j_t})$, where $h$ is a positive, $t$-variate function which is bounded away from zero and infinity, and $\tau(X)^2$ can be represented in the same way. (17)

Assumption (a) can be relaxed to a mixing condition, and (b) can be removed by using the estimated $\psi_j$s rather than their true forms. The latter approach requires a regularity condition on the spacings of the eigenvalues $\theta_j$, and in that setting a complex technical argument, similar to those given by Hall and Hosseini-Nasab (2008), is needed. Condition (c) can be relaxed by noting that if $\sigma(\cdot)$ is sufficiently regular, and if the scores $X_j$ are independent, then $\sigma(X)$ can be approximated by a sequence of functions $\sigma_t(X_1, \ldots, X_t)$, for $t \geq 1$, where $\sigma(X) - \sigma_t(X_1, \ldots, X_t)$ converges to zero as $t \to \infty$, with a similar constraint imposed on $\tau(X)$.

Recall that the pair $(X, Y)$ is generated by the model at (2), where the error, $\epsilon$, has zero mean, and we wish to estimate $\mu(x) = \alpha + \int_{\mathcal{I}} \beta \, x$ for a particular function $x$.

Our estimator, which is equivalent to that given at (14) with $w = \tau(X_{[i]})^{-2}$, is defined by $\bar{\mu}_w(x) = \bar{Y}_w + \sum_{j \leq r} \hat{b}_j (x_j - \bar{X}_{j,w})$, where $\bar{Y}_w = \{\sum_i \tau(X_{[i]})^{-2}\}^{-1} \sum_i \tau(X_{[i]})^{-2} Y_i$ and $\bar{X}_{j,w} = \{\sum_i \tau(X_{[i]})^{-2}\}^{-1} \sum_i \tau(X_{[i]})^{-2} X_{ij}$ and $\hat{b}_1, \hat{b}_2, \ldots$ are chosen to minimise

$$\sum_{i=1}^{n} \left\{ Y_i - \bar{Y}_w - \sum_{j=1}^{r} b_j (X_{ij} - \bar{X}_{j,w}) \right\}^2 \frac{1}{\tau(X_{[i]})^2} . \tag{18}$$

Since we centre the principal component scores $X_{ij}$ at their respective means $\bar{X}_{j,w}$ then we may, and do, assume without loss of generality that $E\{X \tau(X)^{-2}\} = 0$.

The eigenfunctions $\psi_j$ and eigenvalues $\theta_j$ are defined by (6), and, in addition to (17), we assume that:

$$\text{the principal components } \int_{\mathcal{I}} X \psi_j \text{ are independent}, \tag{19}$$

$$\sum_{j=1}^{\infty} \theta_j^{-1} x_j^2 = \infty, \quad \int_{\mathcal{I}} \beta^2 < \infty, \tag{20}$$

$$r = r(n) \to \infty \text{ as } n \to \infty, \text{ and } r = O(n^{-\eta + (1/2)}) \text{ for some } \eta \in (0, \tfrac{1}{2}), \tag{21}$$

$$E\|X\|^k < \infty, \quad \sup_{j \geq 1} \theta_j^{-k} E\left(\int_{\mathcal{I}} X \psi_j\right)^{2k} < \infty \quad \text{for each integer } k \geq 1. \tag{22}$$

Write $\mathrm{AMSE}\{\tilde{\mu}_w(x) - \mu(x)\}$ for the asymptotic mean squared error of the estimator. The following theorem, which is derived in section 5, describes asymptotic properties of this quantity.

**Theorem 3.1.** *If* (17) *and* (19)–(22) *hold then as $n$ and $r$ diverge together,*

$$\mathrm{AMSE}\{\bar{\mu}_w(x) - \mu(x)\} = n^{-1} \frac{E\{\sigma(X)^2 \tau(X)^{-4}\}}{[E\{\tau(X)^{-2}\}]^2} \sum_{j=1}^{r} \theta_j^{-1} x_j^2 + \left(\sum_{j=r+1}^{\infty} b_j x_j\right)^2 . \tag{23}$$

Among the implications that can be drawn from (23) are the following:
(i) If the model, $\tau^2$, for the variance, $\sigma^2$, is essentially correct, i.e. if $\tau$ equals a constant multiple of $\sigma$, then the factor, $\rho^2 \equiv E\{\sigma(Z)^2 \tau(Z)^{-4}\} [E\{\tau(Z)^{-2}\}]^{-2}$, outside the first term in (23), which represents the variance contribution to asymptotic mean squared error, reduces to simply $[E\{\sigma(X)^{-2}\}]^{-1}$; whereas that factor would be simply $E\{\sigma(X)^2\}$ if we were to use unweighted least-squares, i.e. if we were to take $\tau(X)$ to be constant. The fact that, by Jensen's inequality, $[E\{\sigma(X)^{-2}\}]^{-1} \leq E\{\sigma(X)^2\}$, demonstrates the effectiveness of the adaptive approach.
(ii) If the model is essentially incorrect, i.e. if $\tau$ does not equal a constant multiple of $\sigma$, then the estimator remains consistent and enjoys the same convergence rate as before, but with an inflated constant multiplier. More generally, if the variance functional $\sigma^2$ is not constant, and if the model is wrong but approximately correct (in particular, if $\tau(X)$ is sufficiently close to $\sigma(X)$ for sufficiently many values of $X$), then $\rho^2$ is reduced relative to the value it would have if we were to simply take $\tau \equiv 1$. These properties point to the fact that it is not essential to use a nonparametric estimator of variance in order to obtain an improvement in performance over standard least-squares. As noted in section 2.3, for the small to moderate sample sizes commonly encountered with functional data, weighted least-squares based on nonparametric estimators of variance generally perform poorly because the estimated weights introduce too much extra variability.

(iii) The factor $\rho^2$, defined in (i) above, is applied to each and every term in the series $\sum_{j \leq r} \theta_j^{-1} x_j^2$ in (23); it does not reduce in size as $j$ increases. Therefore the advantages of correcting for heteroscedasticity are valid with equal force for arbitrarily large dimension; they do not relate just to low-dimensional aspects of the problem.

(iv) As is to be expected, the effect of weighting has an impact only on the variance contribution to asymptotic mean squared error, not on the bias component. However, even if the problem is finite-dimensional the impact of the variance component persists even in the asymptotic limit, and so there is always something to be gained, in asymptotic terms, by adapting the estimator appropriately to heteroscedasticity.

(v) The first part of (20) determines that the estimator $\widetilde{\mu}_w(x)$ has nonparametric, rather than parametric, convergence rates. It holds if we treat $x$ as a realisation of $X$, and average over all such realisations. In particular, if $x$ is distributed like $X$ then $E(x_j^2) = \theta_j$, and so $E(\sum_{j \geq 1} \theta_j^{-1} x_j^2) = \sum_{j \geq 1} 1 = \infty$, implying that the first part of (20) holds "on average."

## 4 Numerical illustrations

### 4.1 Real data example

We applied our method to Australian rainfall data available at http://www.worldweather.org. The data consist of rainfall measurements at each of 206 Australian weather stations, averaged over the 40 to 100 year periods for which the weather stations had been in use. For any given station we took $Y$ to equal the average yearly rainfall divided by the average number of days on which it had rained, which we refer to as intensity. Our predictor $X(t)$ equalled the rainfall at time $t$, where $t$ represented the fraction of the year that had passed at the time of measurement, and rainfall at that time was averaged over the years for which the station had been operating and was obtained by passing a local polynomial smoother through discrete observations.

The majority of weather stations fall into one of two classes, which respectively comprise most stations in southern parts of the continent, which tend to follow a "European" rainfall pattern where the majority of rain comes in cooler months and summer is relatively dry; and most stations in northern regions, which exhibit a "tropical" pattern where most rain falls in mid to late summer and the cooler months are generally dry. Only a small number of weather stations have more complex rainfall patterns that are not of one of these two types, although some northern stations reflect rainfall southern patterns, and vice versa. These features suggest that most of the data might reasonably be assumed to come from a mixture of two populations. Those populations might produce different error variances in the linear model, leading to heteroscedasticity.

We removed two weather stations, Ipswich and Katherine, which appeared to be outliers, in the sense that the functional linear model explained their rainfall relatively poorly. Then we applied our method to the $n = 204$ remaining stations. To test the method we generated $B = 500$ samples, each of size $n = 102$, by randomly removing half of the 204 stations. For each of the $B$ samples we then applied our method to predict the value of $Y$ for each of the 102 removed stations. Note that, since these were real data, we did not know the true value of the target $\mu$, and so we compared the predicted value with the true value of $Y$. For each of the $B$ samples we calculated the mean squared errors for the 102 predicted stations, that is, for $b = 1, \ldots, B$, we calculated

$$\widetilde{\mathrm{MSE}}_{w,b} = \frac{1}{102} \sum_{i=1}^{102} \{\tilde{\mu}_w(X_{[i]}) - Y_i\}^2, \quad \mathrm{M\check{S}E}_{w,b} = \frac{1}{102} \sum_{i=1}^{102} \{\check{\mu}_w(X_{[i]}) - Y_i\}^2$$

$$\text{and} \quad \mathrm{MSE}_b = \frac{1}{102} \sum_{i=1}^{102} \{\hat{\mu}(X_{[i]}) - Y_i\}^2.$$

For each of the $B$ samples, we also calculated the proportion of the 102 predicted stations that were better predicted by the weighted methods, that is, for $b = 1, \ldots, B$, we calculated $\tilde{p}_{w,b} = \#\{[\tilde{\mu}_w(X_{[i]}) - Y_i]^2 < [\hat{\mu}(X_{[i]}) - Y_i]^2\}/102$ and $\check{p}_{w,b} = \#\{[\check{\mu}_w(X_{[i]}) - Y_i]^2 < [\hat{\mu}(X_{[i]}) - Y_i]^2\}/102$.

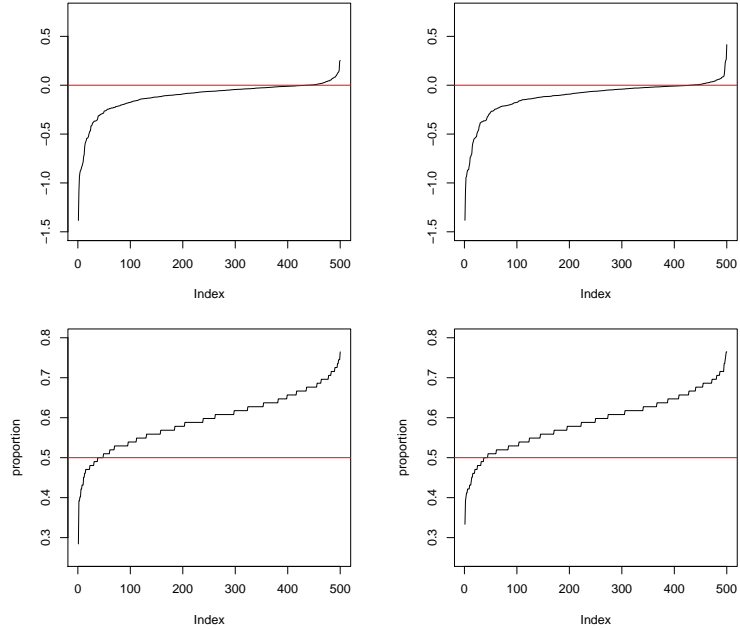

Figure 1: Plot of the 500 ordered values of $\log(\widetilde{\mathrm{MSE}}_{w,b}/\mathrm{MSE}_b)$, top left, $\log(\check{\mathrm{MSE}}_{w,b}/\mathrm{MSE}_b)$, top right, $\tilde{p}_{w,b}$, bottom left, and of $\check{p}_{w,b}$, bottom right. Horizontal lines are for reference only.

In Figure 1 we present graphs of the resulting $B = 500$ ordered values of $\log(\widetilde{\mathrm{MSE}}_{w,b}/\mathrm{MSE}_b)$, of $\log(\check{\mathrm{MSE}}_{w,b}/\mathrm{MSE}_b)$, of $\tilde{p}_{w,b}$ and of $\check{p}_{w,b}$. We see that both weighted methods gave very similar results, and that both strongly bettered the unweighted predictor $\hat{\mu}$: for most of the 500 samples, the MSE of the weighted methods was inferior to that of the unweighted method. Moreover, in more than 90% of the cases, the proportions $\tilde{p}_{w,b}$ and $\check{p}_{w,b}$ were higher than 0.5, meaning that for most of the $B = 500$ samples, more than half of the 102 predicted values were closer to the true $Y_i$ when using the weighted method rather than the unweighted method.

### 4.2 Simulations

Using the same 204 functions $X(t)$ as in section 3.1, we also tested the weighted methods on some generated $Y$ data. The advantage of simulated data is that we know the target $\mu(x) = E(Y|X = x)$, and thus it is easier to assess the quality of the predictors. We generated 204 $Y$-values according to the model at (2), where, we took $\alpha = 0$, $\beta(t) = 0.02 \cdot \sin\{8 - (\pi/20t)\}(1_{\{t \leq 190\}} + 0.5 \cdot 1_{\{t > 190\}})$ and $\epsilon = f(X)U$ where $U \sim U[-3/4, 3/4]$. We tried two models for $f(X)^2$: $(i)$ $f(X)^2 = 0.1 \cdot \{\int \beta(t) X(t)\, dt\}^2$ and $(ii)$ $f(X)^2 = 0.1 \cdot \{\int \beta(t) X(t)\, dt\}^2 + 0.2 \cdot |\int \beta(t) X(t)\, dt|^{1/2}$. Note that the function $\beta$ as defined above is a smoothed and simplified version of the estimated $\beta$ of the real data of section 3.1.

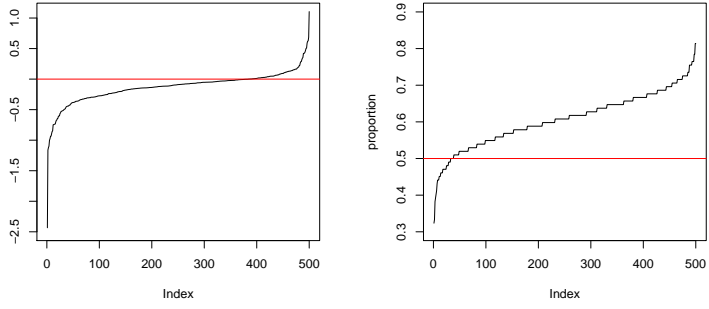

Figure 2: Plot of the 500 ordered values of $\log(\widetilde{\text{MSE}}_{w,b}/\text{MSE}_b)$, left, and of $\tilde{p}_{w,b}$, right, for the generated data with $f(X)^2$ as in $(i)$. Horizontal lines are for reference only.

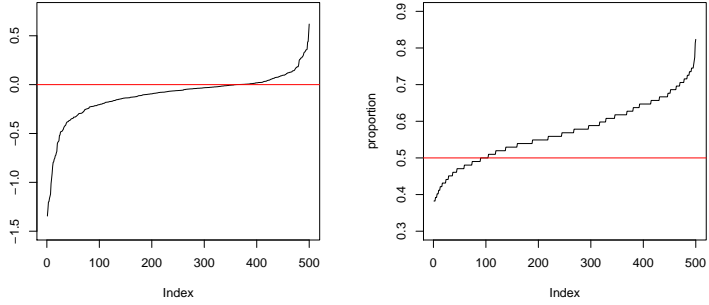

Figure 3: Plot of the 500 ordered values of $\log(\widetilde{\text{MSE}}_{w,b}/\text{MSE}_b)$, left, and of $\tilde{p}_{w,b}$, right, for the generated data with $f(X)^2$ as in $(ii)$. Horizontal lines are for reference only.

We proceeded as in section 3.1 and, by randomly splitting the data $(X_{[i]}, Y_i)$ into two parts, constructing $B = 500$ samples of size $n = 102$, and each time applying the method to the 102 remaining data points. In both cases $(i)$ and $(ii)$ we took the function $g$ at (10) equal to $g(u) = |c_1 u|^{c_2}$. Even though this form is only an approximation to the real $g$ for case $(ii)$, we shall see below that the weighted methods worked well there too.

In this case since we knew the target $\mu$, we replaced $Y_i$ by $\mu(X_{[i]})$ in the definitions of $\widetilde{\text{MSE}}_{w,b}$, $\text{MSE}_b$, $\check{\text{MSE}}_{w,b}$, $\tilde{p}_{w,b}$ and $\check{p}_{w,b}$. Figures 2 and 3 show, respectively, the results for cases $(i)$ and $(ii)$. We show only the predictor $\tilde{\mu}_w$; the results for $\check{\mu}_w$ were similar. The figures illustrate the improvement that can be gained by using a weighted version of the predictor.

# 5 Proof of Theorem

We give the result first in the homoscedastic case, where we take both $\sigma$ and $\tau$ to equal constants, and then we generalise it to the heteroscedastic setting. The model (2) can be written equivalently as $Y = \alpha + \int_{\mathcal{I}} \beta (X - EX) + \epsilon$, for the same function $\beta$ but for a different scalar $\alpha$, which now equals $E(Y)$. We shall work with this model below. The least-squares estimator of $\mu(x)$ is the same as before, but the corresponding estimator of $\alpha$ is now simply $\hat{\alpha} = \bar{Y}$. In particular, using the new model and making assumptions (20) and (22), $\hat{\alpha}$ is root-$n$ consistent for $\alpha$:

$$\hat{\alpha} - \alpha = O_p(n^{-1/2}). \tag{24}$$

The least-squares estimators $\hat{b}_1, \ldots, \hat{b}_r$ are the solutions of

$$\hat{S}\,(\hat{b}_1, \ldots, \hat{b}_r)^{\mathrm{T}} = \hat{s}, \tag{25}$$

where $\hat{S} = (\hat{s}_{j_1 j_2})$ is an $r \times r$ matrix, $\hat{s} = (\hat{s}_1, \ldots, \hat{s}_r)^{\mathrm{T}}$ is an $r$-vector,

$$\hat{s}_{j_1 j_2} = \frac{1}{n} \sum_{i=1}^{n} (X_{i j_1} - \bar{X}_{j_1})(X_{i j_2} - \bar{X}_{j_2}), \quad \hat{s}_j = \frac{1}{n} \sum_{i=1}^{n} (X_{ij} - \bar{X}_j)(Y_i - \bar{Y}), \tag{26}$$

$X_{ij} = \int X_{[i]}\,\psi_j$, $\bar{X}_j = n^{-1}\sum_i X_{ij}$ and $\bar{Y} = n^{-1}\sum_i Y_i$. Without loss of generality, each $E(X_{ij}) = 0$. Put $Z_{ij} = X_{ij}\,\theta_j^{-1/2}$. Then the variables $Z_{ij}$ have zero mean and unit variance, and $Z_{i_1 j_1}$ and $Z_{i_2 j_2}$ are independent for arbitrary $i_1, i_2$ and for $j_1 \neq j_2$ (see (19)). In this notation, $\hat{s}_{j_1 j_2} = (\theta_{j_1}\theta_{j_2})^{1/2}\,\hat{t}_{j_1 j_2}$ and $\hat{s}_j = \theta_j^{1/2}\,\hat{t}_j$, where

$$\begin{aligned}
\hat{t}_{j_1 j_2} &= \frac{1}{n}\sum_{i=1}^{n}(Z_{i j_1} - \bar{Z}_{j_1})(Z_{i j_2} - \bar{Z}_{j_2}), \\
\hat{t}_j &= \frac{1}{n}\sum_{i=1}^{n}(Z_{ij} - \bar{Z}_j)(Y_i - \bar{Y}) = \frac{1}{n}\sum_{i=1}^{n}(Z_{ij} - \bar{Z}_j)\left\{\int_{\mathcal{I}}\beta(X_{[i]} - \bar{X}) + \epsilon_i - \bar{\epsilon}\right\} \\
&= \sum_{k=1}^{\infty} b_k\,\theta_k^{1/2}\,\hat{t}_{jk} + \hat{u}_j, \quad \hat{u}_j = \frac{1}{n}\sum_{i=1}^{n}(Z_{ij} - \bar{Z}_j)(\epsilon_i - \bar{\epsilon}),
\end{aligned}$$

$\bar{Z}_j = n^{-1}\sum_i Z_{ij}$ and $\bar{\epsilon} = n^{-1}\sum_i \epsilon_i$. Define too $\hat{v}_j = \sum_{k \geq r+1} b_k\,\theta_k^{1/2}\,\hat{t}_{jk}$, and put $\widehat{T} = (\hat{t}_{j_1 j_2})$ and $D = \mathrm{diag}(\theta_1^{1/2}, \ldots, \theta_r^{1/2})$, denoting $r \times r$ matrices, and $\hat{u} = (\hat{u}_1, \ldots, \hat{u}_r)^{\mathrm{T}}$, $\hat{v} = (\hat{v}_1, \ldots, \hat{v}_r)^{\mathrm{T}}$, $\hat{b} = (\hat{b}_1, \ldots, \hat{b}_r)^{\mathrm{T}}$ and $b = (b_1, \ldots, b_r)^{\mathrm{T}}$, representing $r \times 1$ vectors. In this notation, (25) is equivalent to $\widehat{T}D\,(\hat{b} - b) = \hat{u} + \hat{v}$. Define $\|A\|^2 = \sum_{j_1}\sum_{j_2} a_{j_1 j_2}^2$. We shall show shortly that $\|A\| \to 0$ in probability as $n \to \infty$; see the paragraph containing (29). Therefore the probability that $\widehat{T} = I + A$ is invertible converges to 1. When $\widehat{T}$ is invertible,

$$\bar{\mu}_w(x) - \mu(x) - (\hat{\alpha} - \alpha) + \sum_{j=r+1}^{\infty} b_j\, x_j = \sum_{j=1}^{r} (\hat{b}_j - b_j)\, x_j = \sum_{j=1}^{r} \left\{ (\widehat{T}\, D)^{-1} (\hat{u} + \hat{v}) \right\}_j x_j.  \tag{27}$$

Write $\hat{t}_{j_1 j_2} = \delta_{j_1 j_2} + a_{j_1 j_2}$, where $\delta_{j_1 j_2}$ denotes the Kronecker delta and $A = (a_{j_1 j_2})$ is an $r \times r$ random matrix. Assuming that $\nu \equiv \|A\| \to 0$ in probability,

$$\begin{aligned}
\widehat{T}^{-1} = (I + A)^{-1} &= I - A + \ldots + (-1)^{k-1} A^{k-1} + (-1)^k A^k \left( I - A + A^2 - \ldots \right) \\
&= I - A + \ldots + (-1)^{k-1} A^{k-1} + A^k A_k,
\end{aligned}$$

where $A_k = (a_{k, j_1 j_2})$ denotes a random matrix and $\|A_k\| \leq (1 - \nu)^{-1}$. Therefore, writing $A^k = (a_{j_1 j_2}^{(k)})$, and defining $c_r = \sum_{j \leq r} x_j^2\, \theta_j^{-1}$, we have,

$$\begin{aligned}
&\left( \sum_{j=1}^{r} \left[ D^{-1} \left\{ \widehat{T}^{-1} - I + A - \ldots + (-1)^k A^{k-1} \right\} (\hat{u} + \hat{v}) \right]_j x_j \right)^2 \\
&= \left[ \sum_{j=1}^{r} \left\{ D^{-1} A^k A_k\, (\hat{u} + \hat{v}) \right\}_j x_j \right]^2 \\
&\leq c_r \sum_{j=1}^{r} \left[ \left\{ A^k A_k\, (\hat{u} + \hat{v}) \right\}_j \right]^2 = c_r \sum_{j=1}^{r} \left\{ \sum_{j_1=1}^{r} \sum_{j_2=1}^{r} a_{j j_1}^{(k)}\, a_{k, j_1 j_2}\, (\hat{u} + \hat{v})_{j_2} \right\}^2 \\
&\leq c_r \left\{ \sum_{j=1}^{r} \sum_{j_1=1}^{r} (a_{j j_1}^{(k)})^2 \right\} \sum_{j_1=1}^{r} \left\{ \sum_{j_2=1}^{r} a_{k, j_1 j_2}\, (\hat{u} + \hat{v})_{j_2} \right\}^2 \\
&\leq c_r\, \|A^k\|^2\, \|A_k\|^2 \sum_{j=1}^{r} \left\{ (\hat{u} + \hat{v})_j \right\}^2.
\end{aligned} \tag{28}$$

Assumptions (19) and (22) imply that, for each integer $\ell \geq 1$,

$$\sup_{1 \leq j_1, j_2 \leq r} E\left( a_{j_1 j_2}^{2\ell} \right) = O\left( n^{-\ell} \right). \tag{29}$$

Therefore, $E(\|A^k\|^2) = O\{(r^2/n)^k\}$, and so, since (21) implies that $r/n^{1/2} \to 0$, we have $\nu \to 0$ in probability. The property $r/n^{1/2} \to 0$ also entails $\|A_k\| = O_p(1)$. Furthermore, $E(\hat{u}_j^2) = O(n^{-1})$ uniformly in $j \geq 1$, and if $1 \leq j \leq r$ then $|\hat{v}_j| = |\sum_{k \geq r+1} b_k\, \theta_k^{1/2}\, a_{jk}|$. The latter result, (22), (29) and the fact that $(\sum_j |b_j|\, \theta_j^{1/2})^2 \leq (\sum_j b_j^2)(\sum_j \theta_j) < \infty$ imply that $E(\hat{v}_j^2) = O(n^{-1})$, uniformly in $1 \leq j \leq r$. (Note that, in view of the second part of (20), $\sum_j b_j^2 < \infty$, and by (22), $E\|X\|^2 = \sum_j \theta_j < \infty$.)

Combining these results we deduce that the right hand side of (28) equals

$$O_p\left\{c_r \left(r^2/n\right)^k r\, n^{-1}\right\} = O_p\left(c_r\, r^{2k+1}\, n^{-(k+1)}\right).$$

Hence, by (27) and (28),

$$\bar\mu_w(x) - \mu(x) - (\hat\alpha - \alpha) = \sum_{j=1}^{r} \theta_j^{-1/2} x_j \left[\{I - A + \ldots + (-1)^{k-1} A^{k-1}\}(\hat u + \hat v)\right]_j$$

$$- \sum_{j=r+1}^{\infty} b_j\, x_j + O_p\left(c_r^{1/2}\, r^{k+(1/2)}\, n^{-(k+1)/2}\right). \quad (30)$$

Using the fact that $r^2/n \to 0$ it can be shown by direct calculation that, for each integer $k \geq 1$,

$$E\left[\sum_{j=1}^{r} \theta_j^{-1/2} x_j \left\{A^k(\hat u + \hat v)\right\}_j\right]^2 = o(c_r\, n^{-1}). \quad (31)$$

Taking $k$ arbitrarily large, and using (30), (31) and the fact that $r = O(n^{-\eta+(1/2)})$ for some $\eta > 0$ (see (21)), we deduce that,

$$\bar\mu_w(x) - \mu(x) - (\hat\alpha - \alpha) = V - \sum_{j=r+1}^{\infty} b_j\, x_j + o\left(c_r^{1/2}\, n^{-1/2}\right), \quad (32)$$

where $V = \sum_{j \leq r} \theta_j^{-1/2} x_j (\hat u + \hat v)_j$.

Note too that, since we are addressing the homoscedastic case,

$$E(V^2) = \sum_{j_1=1}^{r}\sum_{j_2=1}^{r} (\theta_{j_1}\theta_{j_2})^{-1/2} x_{j_1} x_{j_2}\left\{\sigma^2 E(\hat t_{j_1 j_2}) + E(\hat v_{j_1}\hat v_{j_2})\right\}. \quad (33)$$

Now, $E(\hat t_{j_1 j_2}) = n^{-1}(1 - n^{-1})\delta_{j_1 j_2}$ and, recalling that each $E(Z_{ij}) = 0$,

$$n\, E(\hat t_{j_1 k_1}\, \hat t_{j_2 k_2}) = E\left\{(Z_{1j_1} - \bar Z_{j_1})(Z_{1k_1} - \bar Z_{k_1})(Z_{1j_2} - \bar Z_{j_2})(Z_{1k_2} - \bar Z_{k_2})\right\}$$

$$= (1 - n^{-1})^2 E(Z_{1j_1} Z_{1k_1} Z_{1j_2} Z_{1k_2})$$

$$= (1 - n^{-1})^2 \delta_{j_1 j_2}\, \delta_{k_1 k_2},$$

using the properties $j_1, j_2 \leq r$ and $k_1, k_2 \geq r+1$, and the fact that the $Z_{ij}$s are independent. Hence,

$$E(\hat v_{j_1}\hat v_{j_2}) = \sum_{k_1=r+1}^{\infty}\sum_{k_2=r+1}^{\infty} b_{k_1} b_{k_2} (\theta_{k_1}\theta_{k_1})^{1/2} E(\hat t_{j_1 k_1}\, \hat t_{j_2 k_2})$$

$$= n^{-1}(1 - n^{-1})^2 \delta_{j_1 j_2}\, d_r, \quad (34)$$

where $d_r = \sum_{k \geq r+1} b_k^2 \theta_k$. Using (33), (34) and the fact that $d_r \to 0$ as $r \to \infty$, we deduce that, as $r$ and $n$ diverge together,

$$E(V^2) = \frac{1}{n} \sum_{j=1}^{r} \theta_j^{-1} x_j^2 \left\{ (1 - n^{-1}) \sigma^2 + (1 - n^{-1})^2 d_r \right\} \sim \sigma^2 c_r n^{-1}. \qquad (35)$$

In view of the first part of (20), $c_r \to \infty$ as $r \to \infty$. Formula (23), but with $[E\{\sigma(X)^{-2}\}]^{-1}$ replaced by $\sigma^2$, follows from this property, (24), (32) and (35).

Next we outline the argument that extends this result to the heteroscedastic setting. First we discuss a version of the theorem in an artificial problem where the error variance is a function of $Z$, say, which is independent of $(X, Y)$ but is observed along with that pair. That is, the model (2) now has the form $Y = \alpha + \int_\mathcal{I} \beta X + \sigma(Z) \delta$, where the perturbation $\delta$ is independent of $X$ and $Z$ and has zero mean and unit variance. The appropriately weighted criterion function is that at (18) but with $\tau(X_{[i]})$ replaced by $\tau(Z_i)$. In this case the proof above is easily re-worked, in particular with the factor $\tau(Z_i)^{-2}$ included in both series at (26) and in subsequent series, to show that the asymptotic mean squared error of $\bar{\mu}_w(x)$ continues to be given by (23) but with $E\{\sigma(X)^2 \tau(X)^{-4}\} [E\{\tau(X)^{-2}\}]^{-2}$ replaced by $E\{\sigma(Z)^2 \tau(Z)^{-4}\} [E\{\tau(Z)^{-2}\}]^{-2}$. To appreciate the origins of this result, note that in the simpler model where $\beta$ vanishes and $Y = \alpha + \sigma(Z) \delta$, the variance of the weighted least-squares estimator of $\alpha$ is exactly $\rho(n)^2 \equiv E[\{\sum_i \sigma(Z_i)^2 \tau(Z_i)^{-4}\} \{\sum_i \tau(Z_i)^{-2}\}^{-2}]$; and, under the assumption in (17) that $\sigma(z)$ is bounded and $\tau(z)$ is bounded away from zero, $\rho(n)^2 \sim n^{-1} E\{\sigma(Z)^2 \tau(Z)^{-4}\} [E\{\tau(Z)^{-2}\}]^{-2}$.

This result continues to hold when $\sigma$ and $\tau$ are functions of $X$ rather than $Z$, and depend on only a finite number of principal component scores. The proof proceeds by noting first that if $\text{var}(\epsilon \mid X) = h(X_{j_1}, \ldots, X_{j_t})$, as in (17), and if we assume that the components with indices $j_1, \ldots, j_t$ are known and therefore do not need to be estimated, then we are in exactly the position addressed in the previous paragraph; we can take $Z$ to be $(X_{j_1}, \ldots, X_{j_t})$ and replace $X$ by the expansion $\sum'_j X_j \psi_j$ where the summation $\sum'_j$ is over only those indices $j$ not included among $j_1, \ldots, j_t$. Moreover, the asymptotic mean squared error formula is unaffected if we eliminate the components corresponding to $j = j_1, \ldots, j_t$, or if we take those components to be known.